# Formation of photoinduced space-charge field during in-bulk domain creation by femtosecond NIR laser irradiation in MgO:LN crystals

I. A. Kipenko[1], D. A. Zorikhin[1], A. R. Akhmatkhanov[1], V. Ya. Shur[1*]

Ural Federal University, Yekaterinburg, Russia

*Corresponding author: vladimir.shur@urfu.ru

**Abstract**

We have studied the domain switching under NIR fs-laser irradiation in MgO:LN single crystals with special attention to the relative positions of the light-induced domains, microtracks and regions with modified refractive index in the vicinity of the focusing point. The optical imaging along X direction of the irradiated sample demonstrated the narrow microtracks and the lens-shaped regions ("lenses") located in the vicinity of the microtracks. The relative positions of light-induced microtracks, domains and lenses were revealed by overlapping of their images. We have found that the domain envelops the microtrack and partially intersects with the lens. The temperature stability of all light-induced objects during annealing was studied. It was shown that the local modification of the refractive index disappeared irreversibly while the microtrack and domain remain unchanged. The obtained local modification of the refractive index has been considered as a result of the photorefractive effect. The disappearance of the lens after annealing is caused by bulk screening of the photoinduced space-charge field by increased bulk conductivity. The generation of photovoltaic field by tightly focused NIR fs-laser irradiation close to the focusing point is reported for the first time to the authors' knowledge. It should be noted that in LN the photovoltaic field is codirectional with spontaneous polarization and thus cannot switch the polarization. However, it is possible in ferroelectrics with the opposite direction of the photovoltaic field and low value of threshold field. The revealed effect can be utilized for creation of 3D nonlinear photonic crystals by in-bulk domain engineering.

## Introduction

Ferroelectric is a material possessing a spontaneous polarization over some temperature range that can be reversed or reoriented by external electric field. Different signs of nonlinear optical coefficient in ferroelectric domains of opposite polarity opens the unique ability to create the nonlinear photonic crystals for realization of various types of nonlinear optical interactions, such as second-harmonic generation [1,2] and optical parametric oscillation [3,4]. The efficiency of these processes is increased drastically by realization of quasi-phasematching effect by creation of the tailored domain pattern with high period reproducibility by domain engineering methods [1,2,5]. The most popular approach to domain engineering consists in electric-field poling using the lithographically defined periodical electrodes [6]. Despite widespread adoption this method does not allow production of 3D domain structures and is limited in crystal thickness.

Alternative emerging approach to domain engineering is the light-only domain switching offering the advantage of eliminating the electrode fabrication and external field application [7–9]. A prominent branch of these methods uses tightly focused irradiation by near-infrared femtosecond lasers (NIR fs-lasers) for domain creation [10,11]. The crystal transparency for NIR irradiation allows to realize multiphoton absorption in the focus point localized in the bulk leading to local irreversible modification of the refractive index [12]. It was shown recently that this method can be used for local domain switching in the crystal bulk, thus enabling 3D domain structures creation [13–16].

The alternative mechanisms of domain switching under the action of NIR fs-laser pulses include: (1) the action of thermoelectric field [16,17] and (2) pyroelectric field [10,11]. The thermoelectric field is generated by high temperature gradient caused by nonuniform nonlinear absorption of fs-laser irradiation in the vicinity of the focal point [16,17]. In relaxor ferroelectrics with micro-size domain structure the action of this field results in formation of a couple of domains with opposite orientation of spontaneous polarization [16]. The non-uniform pyroelectric field is generated due to crystal heating by fs-laser irradiation and subsequent cooling [10]. The recently proposed mechanism is based on the formation of microtrack in the focal point representing the amorphized micro-size nonpolar volume without spontaneous polarization [18]. The enveloped domain is formed under the action of the depolarization field produced by the bound charges located at the microtrack boundary [19]. It should be noted that the dominating mechanism remains unclear and "full understanding of light-induced ferroelectric domain inversion will require extensive studies in the future" [11].

Single crystals of lithium niobate ($LiNbO_3$, LN) family are one of the main materials for nonlinear optics due to their excellent nonlinear-optical properties and domain structure stability [6,20]. MgO doping of these crystals is used for improving the optical damage durability [21,22]. Domain patterning via NIR fs-lasers in LN crystals has been achieved through several approaches.

One-step methods involve NIR fs-laser irradiation in the crystal bulk only [10,23]. It was demonstrated in a nonpolar cut of MgOLN that the thermoelectric field generated by a moving laser spot could be oriented to either write or erase domain patterns, allowing for the creation of 2D matrices and their transformation into 3D structures with nanoscale precision [17]. The periodical domain structure produced by this method allowed to realize second harmonic generation in potassium titanyl phosphate (KTP) and lead magnesium niobate–lead titanate single crystals [24,25].

Two-step methods of light-only domain engineering have also been developed. One approach creates laser-modified microtracks that act as seeds for domain growing from the microtracks towards the Z- surface during a subsequent uniform heating/cooling cycle, [26–28]. Another two-step method, "laser marking," creates modified surface areas that seed domain growth upon further local heating [29,30].

It should be noted that further development of the light-only domain engineering methods for creation of precise periodical domain structures requires deeper understanding of the localization of the created domains with respect to the focusing point and the formed microtrack. Moreover, it becomes important now to study not only the process in the vicinity of the focusing point, but also the light induced effects far from the focusing point, as they can influence the further forward growth of the formed domain.

In this work we have studied the domain switching under NIR fs-laser irradiation in MgO:LN with special attention to the relative positions of the light-induced domains, microtracks and any regions with changes of the optical properties in the vicinity of the focusing point. We have revealed the appearance of the regions with modified refractive index in the vicinity of the microtrack. We propose that the local modification of the refractive index is due to the photorefractive effect.

**Experimental technique**

A 3-inch Z-cut double-side optical-quality-polished single domain wafer of lithium niobate doped with 5% MgO (Jiangxi Unicrystal Technology, China) was cut into rectangular parallelepiped samples with sizes of 1 × 1 × 10 mm³, with the long side oriented parallel to the Y crystallographic axis. After cutting, the YZ sides of the samples were polished to optical quality.

A femtosecond regenerative amplifier (TETA-10, Avesta Project, Russia) based on an Yb-crystal solid-state laser was employed for local irradiation of the samples. The parameters of the NIR fs-laser irradiation were as follows: wavelength 1030 nm, repetition rate 100 kHz, pulse duration 240 fs, and pulse energy ranging from 2 to 12 µJ. The samples were mounted on a 3D motorized stage (horizontal stage XY5050 combined with vertical stage KA050-Z, Zolix Instruments, China).

Laser irradiation was directed along the Z crystal axis and focused into the bulk of the sample at a depth of 500 µm from the Z- polar surface using a 50× objective LMH-50X-1064 (Thorlabs, USA) with NA = 0.65. The focusing depth was determined taking into account the refractive index of MgO:LN at the 1030 nm wavelength. The irradiation points, arranged in series of three with periods of 100 and 200 µm, were exposed at a fixed depth without laser spot motion (Fig. 1). The number of laser pulses varied from 1 to 1024 pulses, and the delay between successive exposures ranged from 1 to 30 s.

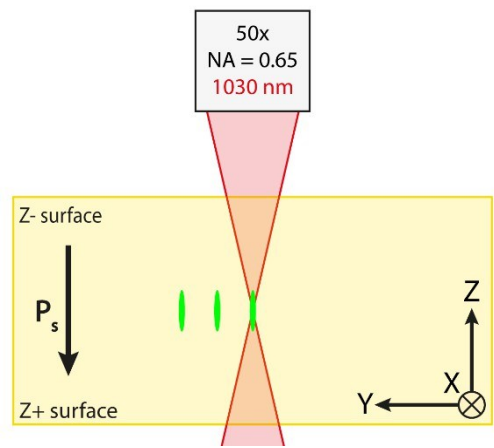

Figure 1 – Scheme of sample irradiation by tightly focused fs-laser pulses.

Several experimental methods were employed for imaging of the formed microtracks and domains. Phase contrast optical microscopy (BX-61, Olympus, Japan) was employed for imaging of the microtracks. The measurements were performed in transmission mode, with the light propagating along either Z or X crystallographic axis to enable imaging of microtracks from the top and the side, respectively.

The domains and microtracks in the bulk were imaged using second-harmonic generation microscopy (SHGM). A custom setup based on Ntegra Spectra Raman microscope (NT-MDT, Russia) with an Yb fiber laser (1064 nm, 5 MHz repetition rate, 2 ns pulse duration) as a radiation source was used. The radiation was focused using a 100× PLAN objective (NA = 0.55). Scanning the samples in the XY cross-section and filtering the light at the SHG wavelength yielded sets of two-dimensional images at various depths, which were subsequently reconstructed into three-dimensional images of microtracks and domains.

For studying the thermal stability of the modifications created by fs-laser irradiation the sample was annealed with Linkam THMS600 thermal-cooling stage (Linkam Scientific Instruments Ltd., UK) and imaged using Linkam Imaging Station with 10× objective PlanC N (Olympus Corporation, Japan) (NA=0.25) and FLIR CMOS camera (FLIR Integrated Imaging Solutions Inc., Canada) with 1 Hz frame rate. Annealing profile consisted of three steps: 1) heating from room temperature to 150 ºC, 2) annealing at 150 ºC for 10 minutes, 3) cooling to room temperature. Heating/cooling rates were 3 ºC/min.

**Results**

The traditional optical observation of the irradiated sample along the polar (Z) axis allows us to reveal the bright spots corresponding to microtracks (Fig. 2a). Their diameter increases with the pulse number. Moreover, the additional ring contrast around the spots appears for pulse number above two (Fig. 2a). The optical imaging of the same sample along X crystallographic direction (YZ projection) allows to reveal the lens-shaped regions ("lens") located in the vicinity of the narrow microtrack closer to the irradiated surface (Fig. 2b,c).

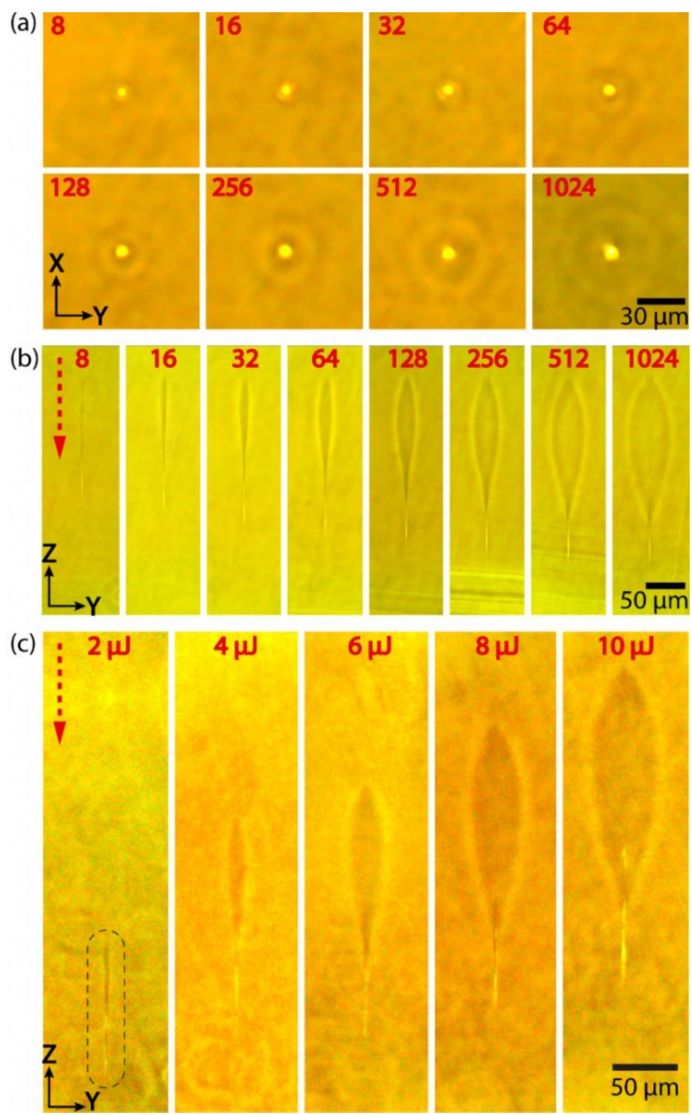

Figure 2 – Optical phase contrast microscopy images of the sample after NIR fs-laser irradiation: imaging along (a) Z axis and (b) X axis for areas irradiated by various number of pulses with 10 µJ pulse energy; (c) imaging along X axis for areas irradiated by various pulse energy with 1000 pulses. Irradiation direction is marked by red dashed arrows at (b) and (c).

SHGM was used for imaging of the domains formed in the sample as a result of fs-laser irradiation. The plotted view along X axis demonstrates the spindle-like domain shape (Fig. 3a). The overlapping of the obtained optical and SHGM images allows us to reveal the relative positions of microtracks, domains and lenses created by laser irradiation (Fig. 3b,c). It is seen that the domain envelops the microtrack. Similar results have been published earlier [31].

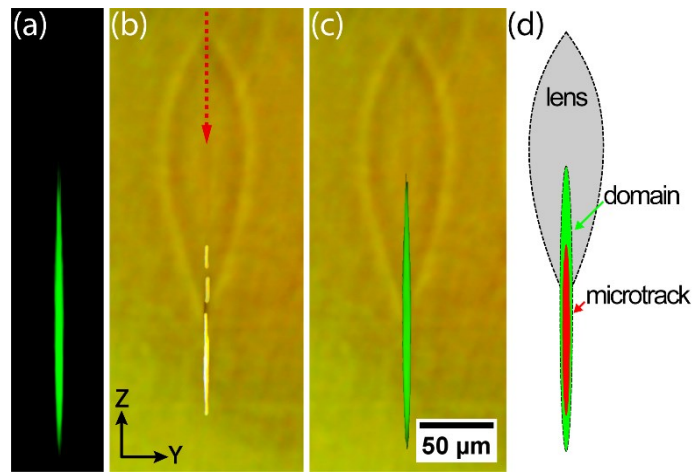

Figure 3 – Optical microscopy and SHGM images of the sample after NIR fs-laser irradiation with 1000 pulses and 10 µJ pulse energy: (a) SHGM domain image; (b) overlapping of optical images of microtrack and lens; (c) overlapping of lens optical image and SHGM domain image. View along X axis. Irradiation direction is marked by red dashed arrow at (b).

It is necessary to point out that the domain and the microtrack only partially intersect with the lens (Fig. 3b,c). Moreover, there is no relation between the shapes of the domain and the lens (Fig. 3c). The lens length is comparable with the length of the domain whereas its width is essentially larger than the domain's one (Fig. 3c).

The temperature stability of all objects created by laser irradiation was studied by optical observation during annealing. The sample was heated to 150°C at rate 3°C/min and annealed at this temperature for 10 min. It was shown that the lens contrast disappeared irreversibly during annealing (Fig. 4). In contrast the microtrack and domain remain unchanged after temperature treatment (Fig. 5). It was found also that the additional bright spot appears during heating at the end of the microtrack located further from the irradiated surface (Fig. 4). This spot disappears during annealing (Fig. 4).

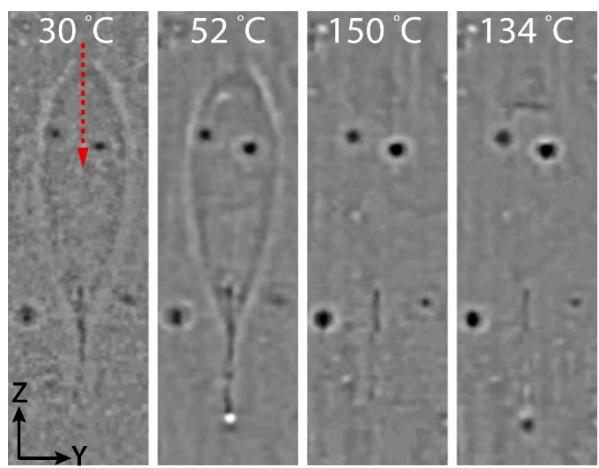

Figure 4 – Images at various temperatures during annealing at 150°C of the sample after NIR fs-irradiation with 1000 pulses and 12 µJ pulse energy. View along X axis. Cross-polarized optical microscopy. Irradiation direction is marked by red dashed arrow.

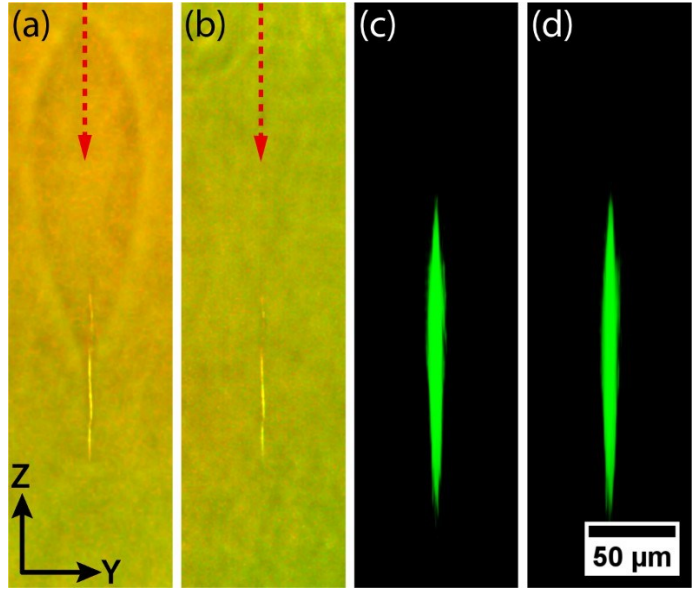

Figure 5 – Images of the sample after NIR fs-irradiation with 1000 pulses and 12 µJ pulse energy (a), (c) before and (b), (d) after annealing. (a), (b) Overlapping of optical microscopy images of lens and microtrack. (c), (d) SHGM domain images. View along X axis. Irradiation direction is marked by red dashed arrows at (a) and (b).

We have studied the dependence of the sizes of all objects created by laser irradiation on the irradiation parameters. The lens sizes increase with pulse energy and number (Figs. 2b,c and 6). The dependence of lens length (*l*) and width (*w*) on pulse energy (*E*) was fitted by the following formulas (Fig. 6c,d):

$$l(E) = l_0 + k_l (E - E_{th}) \qquad (1)$$

$$w(E) = w_0 + k_w (E - E_{th}) \qquad (2)$$

where $l_0 = 61$ µm and $w_0 = 5$ µm are the minimal experimentally obtained length and width for irradiation energy $E_{th} = 2$ µJ, $k_l$ and $k_w$ are the coefficients. The best fit values $k_l = 17$ µm/µJ and $k_w = 6$ µm/µJ.

It should be noted that for the highest used energy $E = 10$ µJ and pulse number $N = 1024$ the lens length and width reached the values of 180 µm and 60 µm, respectively.

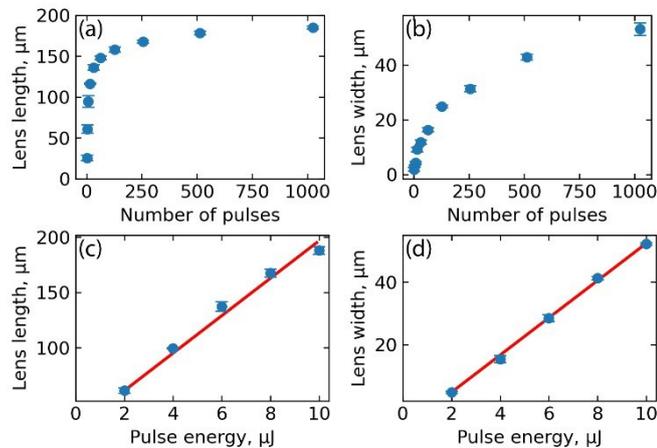

Figure 6 – The lens (a), (c) length and (b), (d) width dependences on (a), (b) number of pulses for 10 µJ energy and (c), (d) pulse energy for irradiation by 1000 pulses.

The length of the microtrack formed as a result of single pulse irradiation with energy 10 µJ is 60 µm. It increases with pulse number and reaches 120 µm for 1024 pulses (Fig. 7a). The minimal observed microtrack length was 55 µm for pulse number 1024 and pulse energy 2 µJ

(Fig. 7b). It increased with pulse energy and reached 100 μm for 10 μJ (Fig. 7b). The microtrack width was not analyzed because its value was always below the optical resolution.

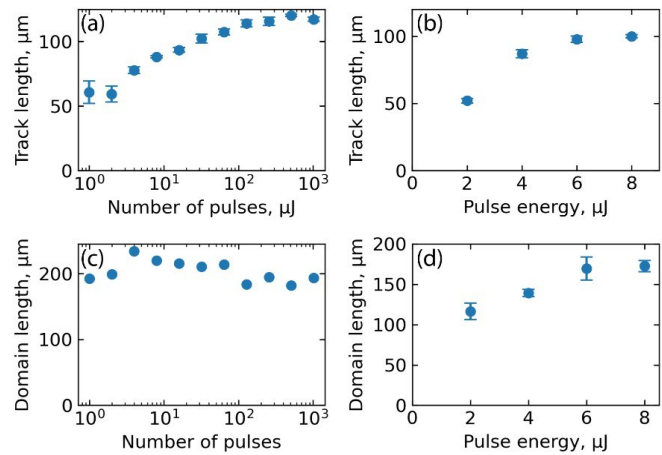

Figure 7 – Dependences of (a), (b) microtrack length and (c), (d) domain length on (a), (c) number of pulses and (b), (d) pulse energy.

The domain is always at least twice as long as the microtrack. The domain length doesn't depend on the pulse number and increases with pulse energy (Fig. 7c,d). The domain width was not analyzed because its value was always below the SHGM resolution.

**Discussion**

The obtained local modification of the refractive index can be considered as a result of the photorefractive effect. The carriers ejected from the deep traps in polar direction as a result of multiphoton absorption generate the photogalvanic current [32]. The redistribution and trapping of photoexcited charges lead to creation of the localized space-charge, which generates a photovoltaic field that modifies the refractive index via the linear electrooptic effect [33]. It should be noted that the photovoltaic field generation has been studied intensively in nominally pure and Fe doped lithium niobate crystals during CW visible light irradiation [34,35]. However, the generation of photovoltaic field as a result of tightly focused irradiation of NIR fs-laser in the vicinity of the focusing point is reported for the first time to the authors' knowledge. While MgO doping of lithium niobate is used to suppress the photovoltaic effect, it looks like that light intensity in the vicinity of the focusing point is high enough for creation of the photogalvanic current induced by multiphoton absorption.

In terms of the proposed model the disappearance of the modification of the refractive index during annealing is caused by bulk screening of the photoinduced space-charge field by increased bulk conductivity [36]. The location of the lens in front of the microtrack closer to the irradiated surface can be attributed to the known effect of the reflection of NIR fs laser irradiation from the plasma generated in the focusing point [37].

The appearance of the bright spot at the end of the microtrack during sample heating can be attributed to local change of the refractive index under the action of the partially screened depolarization field generated by bound charges localized at the boundary of the microtrack. We suppose that the microtrack is the amorphized volume elongated along the polar direction [18]. In this case the maximal density of the bound charges is localized at the microtrack ends. At room temperature the depolarization field produced by bound charges is completely compensated by bulk screening. The decrease of spontaneous polarization during heating leads to appearance of the pyroelectric field due to retardation of the bulk screening of the temperature induced change of the depolarization field. The spot disappears during annealing at constant elevated temperature due to decreasing of the pyroelectric field by effective screening.

It should be noted that in LN single crystals the photovoltaic field is codirectional with spontaneous polarization and thus cannot switch the polarization [34,35]. However, in ferroelectrics with the opposite direction of the photovoltaic field and low value of threshold field [38] the polarization switching in the lens region is possible.

**Conclusion**

We have studied the domain switching under NIR fs-laser irradiation in MgO:LN single crystals with special attention to the relative positions of the light-induced domains, microtracks and regions with modified refractive index in the vicinity of the focusing point. The optical imaging of the irradiated sample along the polar axis revealed the bright spots corresponding to microtracks surrounded by rings with modified refractive index. The optical imaging of the sample along X crystallographic direction demonstrated the narrow microtracks and the lens-shaped regions ("lenses") located in the vicinity of the microtracks closer to the irradiated surface. The overlapping of the obtained optical and SHGM domain images revealed the relative positions of all light-induced objects: microtracks, domains and lenses. We have found that the domain envelops the microtrack and partially intersects with the lens. The lens length is comparable with the length of the domain whereas its width is essentially larger than the domain's one. It was shown that the lens sizes (length and width) increase linearly with fs-pulse energy.

The temperature stability of all light-induced objects was studied by optical observation during annealing. It was shown that the local modification of the refractive index disappeared irreversibly while the microtrack and domain remain unchanged after temperature treatment.

The obtained local modification of the refractive index has been considered as a result of the photorefractive effect due to multiphoton absorption in the vicinity of the microtrack closer to the irradiated surface. The location of the lens in front of the microtrack closer to the irradiated surface has been attributed to the known effect of the reflection of NIR fs laser irradiation from the plasma generated in the focusing point. The disappearance of the lens after annealing is caused by bulk screening of the photoinduced space-charge field by increased bulk conductivity. The generation of photovoltaic field has been studied previously in nominally pure and Fe doped lithium niobate crystals during CW visible light irradiation. However, the generation of photovoltaic field by tightly focused NIR fs-laser irradiation close to the focusing point is reported for the first time to the authors' knowledge. It should be noted that in LN single crystals the photovoltaic field is codirectional with spontaneous polarization and thus can not switch the polarization. However, the polarization switching in the lens region is possible in ferroelectrics with the opposite direction of the photovoltaic field and low value of threshold field. The revealed photoinduced space-charge field can be utilized for creation of 3D nonlinear photonic crystals by in-bulk domain engineering.

**Acknowledgements**